\documentclass[preprint,showpacs,preprintnumbers,amsmath,centertags,amssymb]{revtex4-1}

\usepackage[dvips]{graphicx}
\usepackage[dvips]{color}
\usepackage{dcolumn}
\usepackage{bm}
\usepackage{textcomp}
\usepackage{amsmath}

\begin{document}

\title{Propagation of acoustic surface waves on a phononic surface investigated by transient reflecting grating spectroscopy}

\author{I. Malfanti$^{1,2}$, A. Taschin$^{1,3}$, P. Bartolini$^{1}$, B. Bonello$^{4}$, and R. Torre$^{1,2,\ast}$}

\affiliation{
$^1$European Lab. for Non-Linear Spectroscopy (LENS), Univ. di Firenze,
{\em Via N. Carrara 1, I-50019 Sesto Fiorentino, Firenze, Italy.}\\
$^2$Dip. di Fisica ed Astronomia, Univ. di Firenze,
{\em Via Sansone 1, I-50019 Sesto Fiorentino, Firenze, Italy.}\\
$^3$Istituto Nazionale di Ottica (INO)- CNR, 
{\em Largo E. Fermi 6, I-50125, Firenze, Italy.}
$^4$Institute de NanoSciences de Paris (INSP),
{\em 140 rue de Lourmel, 75015 Paris, France.}\\
$^\ast$ {\em corresponding author: \textsf{torre@lens.unifi.it}}}

\date{\today}

\begin{abstract}
We present a study of surface acoustic waves (SAW) propagation on
a 1D phononic surface (PS) by mean of an heterodyne-detected
transient reflecting grating experiment. We excited and detected
coherent stationary SAWs characterized by variable wave-vectors.
The measured SAW frequencies enables the characterization of the
band diagram of this PS sample beyond the first Brillouin zone
(BZ). Four different SAW frequencies have been revealed, whose band diagram show articulated dispersion phenomena. In order to address the nature of the investigated SAWs, the experimental results are compared with a numerical simulation of elastic modes based on a finite element
model. The observed SAWs are addressed to four Bloch waves characterized by different
frequencies and surface energy localization. Moreover, we measured the SAW propagation on a flat non-phononic part of the sample surface and compared it with results from the PS. 
\end{abstract}

 \maketitle

\section{INTRODUCTION}
The study of propagation of surface acoustic waves (SAW) has relevant interest in both fundamental
research and technological applications, from geophysical studies on seismic waves to the
realization of electro-acoustic devices. Starting with the pioneering works of Lord
Rayleigh~\cite{Rayleigh_1885,Rayleigh_1888} a large number of experimental and theoretical studies
have been published on SAW science~\cite{Landau_59,Auld_73,Victorov_81,Oliner_78}. 

During the last twenty years, there was a great deal of interest for periodic elastic structures,
which are called since that time phononic crystals, by analogy with their optical counterpart the
photonic crystals. The propagation of acoustic waves in phononic crystals shows many peculiar
phenomena that open the possibility for the realization of acoustic metamaterials (for a review
see~\cite{Lu_09}). A clear example is the creation of phononic band gaps\cite{Kushwaha_94} (i.e a
frequency intervals over which the propagation of sound is forbidden), that enables a unique
control on the propagation of sound~\cite{Khelif_04}. Other phenomena are connected with the
acoustic processes characterized by frequencies and wave-vectors at the band edge, where the
folding and bending of the acoustic bands take place. Relevant examples are the negative refraction
phenomena~\cite{Feng_06,Cummer_07} and the focusing processes~\cite{Yang_04,Zhang_09}.

The presence of artificial periodic elastic structure in a material produces relevant variations
even on the nature and proprieties of SAWs. Just a simple periodic modulation of the surface
profile is able to introduce new and interesting modifications of
SAWs~\cite{Auld_76,Glass_81,Gulyaev_98}. The phononic surface phenomena were already an issue in the
technological application of SAWs in the IDTs\cite{Oliner_78}, but recently they met a renewed
interest thanks to the boost of phononic crystal
studies~\cite{Laude_06,Zhang_06,Olsson_09,Bonello_10}. At present, the research on acoustic
proprieties of phononic surfaces (PS) aimed at both the fundamental aspects and the applicative
ones. The complete definition of the PS elastic properties remains to be achieved and the new
properties emerging from the periodic arrangement of the surface itself could develop a new
generation of SAW devices.

A simple 1D PS can be made by engraving parallel grooves on a homogeneous
substrate~\cite{Auld_76,Glass_81,Glass_83}. Contrary, to a flat surface, where only one
non-dispersive Rayleigh mode can exist, in a corrugated surface new modes characterized by
different polarization, penetration and dispersion characteristics appear. Moreover, PS phenomena
manifest when the SAW wavelengths are comparable to the depth of the grooves and they approach the
spatial periodicity. New effects such as opening of band gaps, folding of some branches in the
Brillouin zone, dispersion of SAWs occur~\cite{Zhang_06,Laude_06,Maznev_08}. Yet, the SAWs
characterized by wave-vectors close to the band edge show non-propagative
nature~\cite{Laude_06,Maznev_09}. In fact, at the Brillouin zone (BZ) limit, the SAWs approach the
zero group velocity so that their elastic energy would not propagate remaining spatially localized.
In spite of the great amount of experimental and theoretical studies, a complete and detailed
description of these new acoustic modes and phenomena remains an open problem.

The investigation of SAWs can be performed by different experimental techniques~\cite{Auld_73},
recently the techniques based on laser pulses have proved to be particularly suitable to study the
high frequency SAWs~\cite{Hess_02}. Especially, the transient reflecting grating (TRG)
techniques~\cite{Kasinski_88,Fishman_91,Duggal_92,Sawada_95} enable the excitation and probing of
coherent SAWs measuring their dynamics in the direct time domain. In these experiments two temporally
overlapped picosecond laser pulses are crossed at the surface, the resulting interference pattern
excites two counter propagating SAWs by thermoelastic excitation. The temporal evolution of the
induced surface transient grating is probed by the diffraction of a separate laser beam. The SAW
wave-vector is experimentally fixed by the exciting grating period enabling the scan of the
acoustic band diagram.

Recently mainly two types 1D PS samples have been investigated by pulsed laser spectroscopy. A
first type is realized depositing a surface periodic structure made by parallel strips of metal and
amorphous dielectric matter on a solid substrate~\cite{Antonelli_02,Profunser_06} whereas the second
is made of deep parallel grooves engraved on the homogeneous substrate covered by an uniform metal
thin film~\cite{Dhar_00}. Both these samples have been also investigated by TRG
techniques~\cite{Dhar_00,Maznev_08,Maznev_09}. In these experiments the presence of different SAW
modes, band gaps and long living SAWs have been detected but the full characterization of the
acoustic waves is missed because of some experimental limitations. So, the complete understanding of
the Bloch waves on PS covering the full wave-vector range, inclusive of band edge part, remains
to be achieved.

In this work we report the study of surface waves propagating in a 1D PS made of a grating of
grooves engraved on the surface of an amorphous silica substrate coated with an uniform gold thin
film. The experimental study has been performed by a heterodyne detected transient grating
technique realized in reflection geometry. In particular, in this work we have compared the SAW
band diagram of unpatterned region of the sample with the region presenting the 1D PS structure.
The effect of the phononic structure on the SAW propagation has been observed and discussed. The
experimental data have been compared to those obtained by the simulation realized with the finite
element analysis performed by a commercial software.

\section{EXPERIMENT AND SIMULATION}\label{experiment&sample}

\begin{figure}
    \includegraphics[scale=0.7]{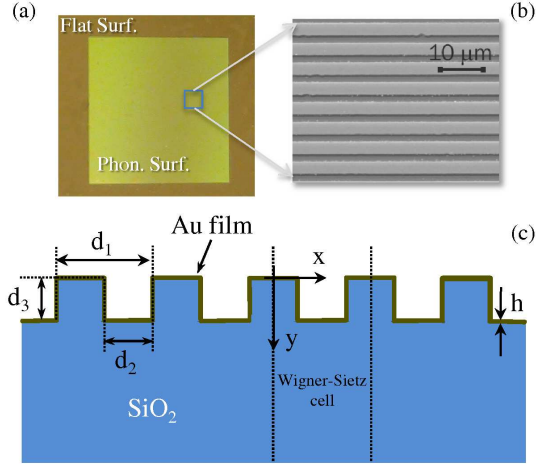}
    \caption{(a) Macroscopic image of the sample. Two regions are identified: the Flat Surface (FS) and the Phononic Surface (PS). In (b) a scanning electron microscopy image of the PS area and in (c) a scheme corresponding to a cross section of the sample perpendicular to the grooving, $d_1=5~\mu$m, $d_2/d_1=52\%$, $d_3=0.860~\mu$m, and $h=0.130~\mu$m}
    \label{sample}
\end{figure}

We characterized a sample made by a fused silica plate where a 1D PS structure has been realized.
An image of the sample is shown in Fig.~\ref{sample}(a). Two distinct area are clearly distinguishable, a flat surface (FS) region and a grooved one ($5\times 5$ mm) that is the phononic surface (PS) part. During this work we have experimentally characterized both regions. The grooving
displayed in Fig.~\ref{sample}(b) is obtained by a photo-lithographic procedure. The 1D surface
pattern has been impressed on a layer of photo-resist, coated on the glass surface, using a optical
mask. The reactive ion etching enables to hollow the square-wave pattern on the glass surface. The
remaining photoresist is then removed with acetone. Finally, in order to excite surface acoustic
waves, a thin gold film is deposited on both the FS and the PS by evaporation. In Fig.~\ref{sample}(c) a
schematic view of the grooving is given. The parameters peculiar of the grating are: the step d$_1=5$  $\mu$m, with a duty cycle of d$_2$/d$_1$=$52\%$, and the depth of the grooves d$_3=0.860$ $\mu$m. The gold film thickness is h=$0.130$ $\mu$m for both the FS and PS regions. The silica plate has a total thickness of 2 mm.
\begin{figure}
    \includegraphics[scale=0.7]{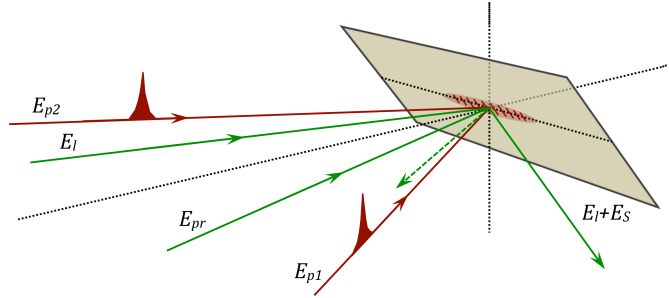}
    \caption{Sketch of the transient grating experiment in the reflection configuration. The two pump pulses $E_{p1}$ e $E_{p2}$ interfere on the sample surface producing a spatially periodic corrugation of the surface. The relaxation of the induced ripple is probed through the measure of the scattered field intensity ($E_{S})$ of the continuous beam $E_{pr}$. Heterodyne detection is achieved by beating $E_{S}$ with the zero order reflection of the local field $E_{l}$ which is exactly collinear to the scattered field.}
    \label{set-up}
\end{figure}
The samples were investigated by means of an heterodyne detected transient reflecting grating
experiment (HD-TRG). TRG is a time resolved spectroscopic technique that allows the
characterization of the relaxation dynamics of reflective media. This experiment is more usually
applied to study transparent materials~\cite{Torre_01,Taschin_06,CapBook_TG2008} in transmission
geometry. In the present TRG experiment (Fig.~\ref{set-up}), two infrared laser pulses, obtained dividing a single pulsed laser beam (pulse duration is $\sim 20$ ps, wavelength 1064 nm and a total intensity at the
sample's surface of a few $\mu$J), are temporally and spatially overlapped on the sample surface
where they interfere producing an impulsive and spatially periodic variation of the temperature due
to metal absorption. The thermoelastic processes modify both the surface
reflectivity~\cite{Thomsen_86} (i.e. induced variations of dielectric tensor via thermo-optic and
elastic-optic coupling tensors) and the surface profile~\cite{Duggal_92,Rogers_00} (i.e. induced
surface ripple via thermal expansion). The first effects are negligible in this experiment and the
signal is dominated by the induced surface ripple~\cite{Rogers_00}. So, the signal turns out to be
only sensitive to the surface vertical displacements (i.e. displacements out from the surface
plane). The thermal expansion generates two counter-propagating mechanical waves with wave-vector
$\mathbf{q}$ defined by the difference of the two pump wave-vectors. Its modulus is
$q=4\pi\sin(\theta_{ex}/2)/\lambda_{ex}$, where $\lambda_{ex}$ and $\theta_{ex}$ are the wavelength
and the incidence angle of the exciting pumps, respectively. A cw laser beam (wavelength 532 nm)
probes the induced grating. The time evolution of the diffracted signal supplies information about
the dynamic of the relaxing TG and, consequently, on the dynamical properties of the analyzed
sample. The experimental details about the laser systems, optical set-up and acquisition procedure
can be found elsewhere~\cite{CapBook_TG2008,Cucini_10}.

We performed a finite element analysis (FEA) of the SAW features in our sample based on a finite element model implemented in the commercial software COMSOL MultiPhysics. This finite element software has the main advantage to be user-friendly, enabling an easy implementation of a FEA of complex samples.
Several possible simulation models could be realized, using the ComsolMultiPhysics platform, relevant to present experiments. A possible one would consist in the definition of a physical model able to describes both the excitation mechanism and the dynamic properties of SAW propagating on a phononic surface, see for example~\cite{Duhring_09}. The numerical solution of such a model would give a deep understanding of the SAW generation and propagation, enabling a more direct comparison with experimental results. Nevertheless this procedure is quite complex and requires an evolute physic and simulation model. A simpler way to characterize the measured SAW on a PS is to use an eigenvalue analysis
for structural mechanical problems~\cite{Nardi_09}, this enables to retrieve the Bloch
eigen-functions and eigen-frequencies that are supported by a given elastic structure. This
approach is already implemented in a COMSOL module software. Once all the geometrical and mechanical
characteristics of the structure are specified, it allows to determine the eigen-frequencies
associated to a specified eigenvector and their associated displacement field. It is therefore
clear that the polarization state at each eigen-frequency can be evaluated.

In the present TRG experimental investigation we need to address the measured SAW frequencies to
the specific Bloch waves and characterize them. So we perform the eigenvalue analysis that enables
the necessary understanding of the elastic properties of the sample without require a more
demanding theoretical and simulation work. The simulations were realized over the Wigner-Seitz cell
(WSC), reported in Fig.~\ref{sample}(c), imposing Bloch conditions at the boundary to the
displacement field \textbf{u}($x$,$y$)=($u$($x$,$y$),$v$($x$,$y$)) (where $u$ and $v$ are
horizontal, $x$, and vertical, $y$, displacement field respectively).

\section{RESULTS}\label{risultati}

The HD-TRG data have been collected at room temperature for a $q$-vector range extending from
$q=0.09~\mu$m$^{-1}$ up to $q=1.04~\mu$m$^{-1}$ which widely covers the first BZ. We performed the
HD-TRG experiments on both the FS and the PS of the sample. In the PS we investigated the SAW with
wave-vectors $q$ parallel or orthogonal to the 1D structure (i.e. parallel or orthogonal to the
grooves). 

\subsection{Flat Surface}\label{hom&par}

\begin{figure}
    \includegraphics[scale=0.60]{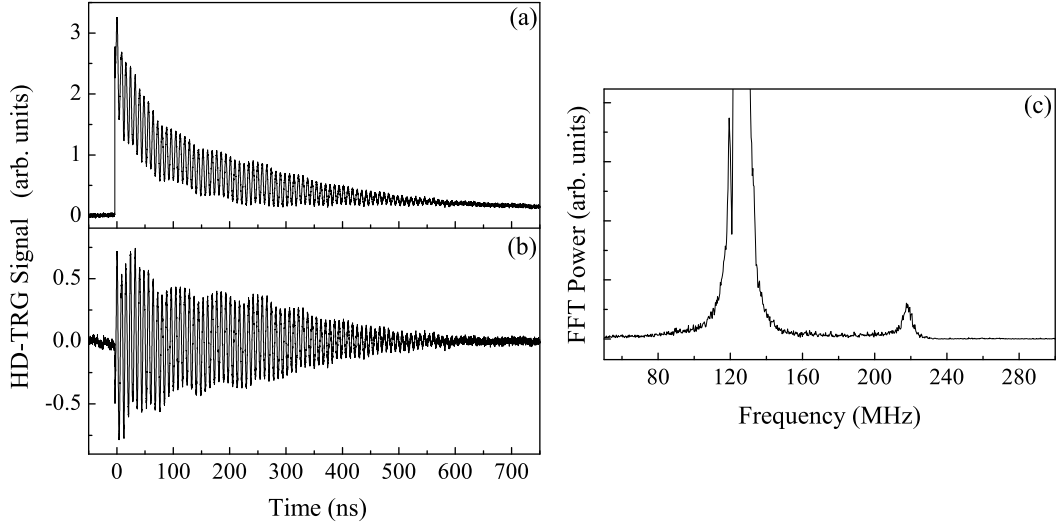}
    \caption{Panel (a): typical time evolution of the HD-TRG signal measured on the FS region at $q=0.25~\mu$m$^{-1}$. The signal is composed of two main contributions: a slowly relaxing term over which a fast oscillating one is clearly observable. Panel (b): signal after the slow term subtraction. Panel (C): FFT of the signal after the slowly relaxing term subtraction. Two acoustic modes are clearly evident.}
    \label{signal&sim}
\end{figure}

\begin{figure}
    \includegraphics[scale=0.35]{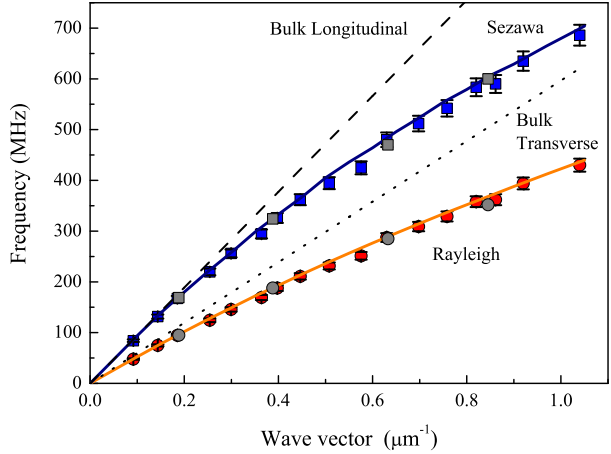}
    \caption{Measured and simulated dispersion of surface modes. Blue square and red circle symbols correspond to experimental measured frequency of surface acoustic modes in FS region, grey squares and circles correspond to data in PS with wave-vectors parallel to the grooves. Red circles (grey circles) represent the Rayleigh mode, blue square (grey squares) correspond to first Sezawa mode (or second mode). Orange and light blue continuous lines correspond to simulated dispersion diagram. Black dotted (dashed) line corresponds to the transverse (longitudinal) mode of the substrate.}
    \label{hom&sim&par_dispersion}
\end{figure}

In the panel (a) of Fig.~\ref{signal&sim} we report a typical HD-TRG data measured on the FS at
$q=0.25~\mu$m$^{-1}$. 
The data show two main contributions: oscillations due to the excited
acoustic surface waves laid upon a slower decay due to the permanent temperature grating relaxing
by thermal diffusion. In order to better infer the acoustic parameters of interest from the signal,
we subtracted the thermal decay, as reported in the panel (b) of Fig.~\ref{signal&sim}, and then
we performed a Fast Fourier Transform (FFT) algorithm of the resulting data. In the panel (c) of
the same figure we show the acoustic spectra content of signal at $q=0.25~\mu$m$^{-1}$, the
presence of two acoustic modes is evident. All HD-TRG data measured on FS showed only the presence
of these two modes. We have then induced waves on the patterned region with wave-vector parallel to
the grooves and, as we expected, the two modes didn't show changes in the relation dispersion (see
Fig.~\ref{hom&sim&par_dispersion} grey circles and square respectively). The induced waves do not
sense the surface structure as long as it propagates parallel to grooves since, being contained in
the sagittal plane, no periodicity is present. That staten, please note that, when in the
following, we will generally refer to the FS data also the case of propagation in the patterned
sample with wave-vector parallel to grooves will be considered. The relative dispersion relations are shown in Fig.~\ref{hom&sim&par_dispersion}. Red ones
correspond to the fundamental mode and blue ones to the second weaker mode. Consistently with the
previous considerations the dependence of the frequency on the wave-vector is not linear for both
the waves.

\subsection{Phononic Surface}

\begin{figure}
    \includegraphics[scale=0.60]{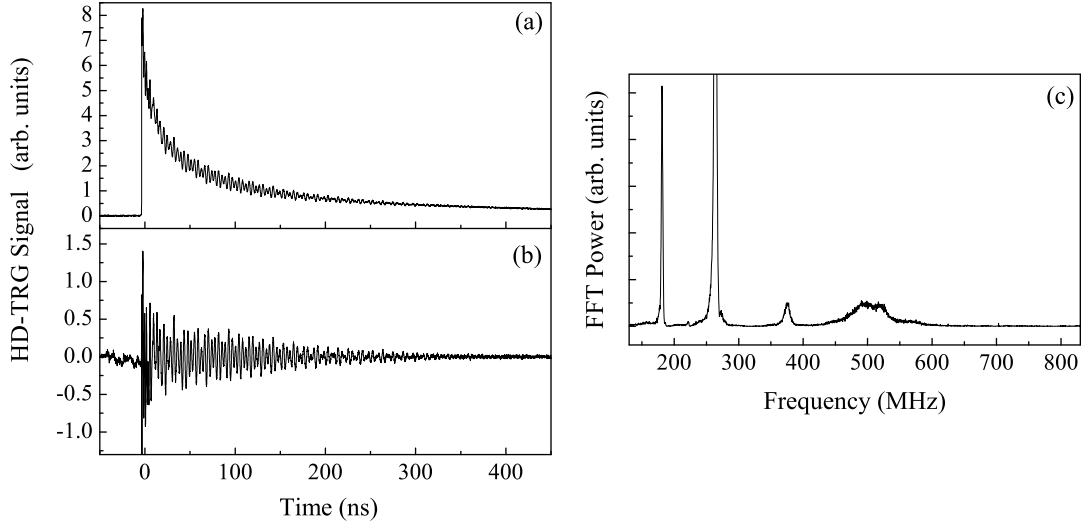}
    \caption{Panel (a): Temporal behaviour of the HD-TRG signal measured on the PS for wave
    vector ($q=0.82~\mu$m$^{-1}$) propagating orthogonal to the grooves. Panel (b): signal after
    the slow term subtraction. Panel (c): FFT of the signal after the slowly relaxing term
    subtraction. Four acoustic modes are clearly evident.}
    \label{data&PS//q}
\end{figure}

\begin{figure*}
    \includegraphics[scale=0.5]{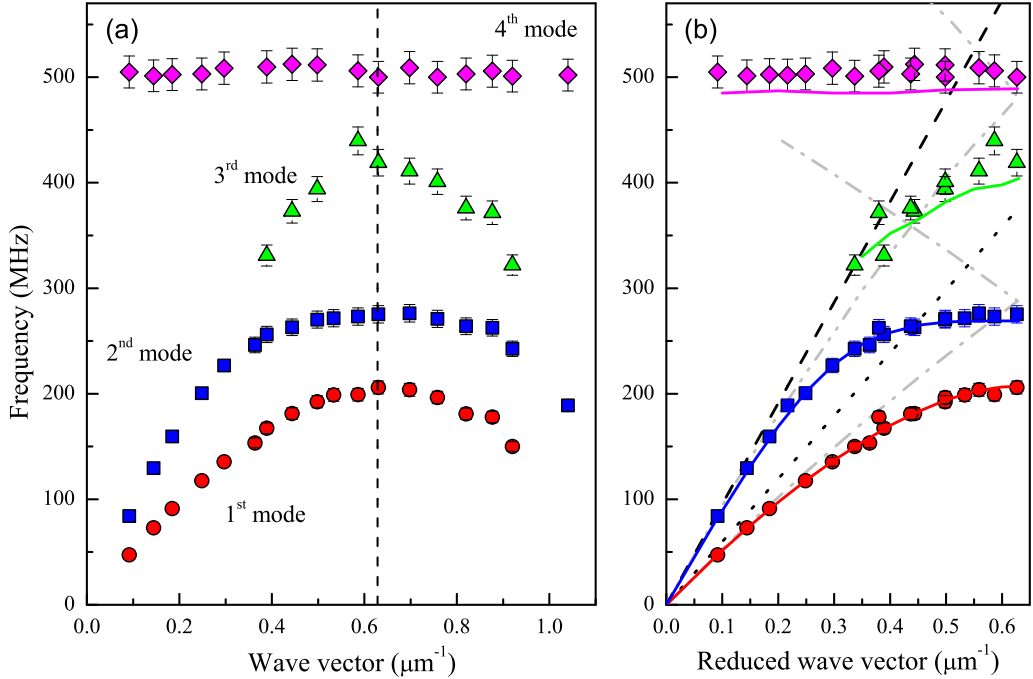}
    \caption{
    (a) Measured dispersion curves on the PS, raw data. Red circles and blue squares
    (we will generally refer to these as first and second mode respectively)
    are analogue to those modes observed in the FS region.
    Green triangles and magenta diamonds are respectively third and fourth mode and do not have any counterpart in the FS region.
    Vertical dashed line is in correspondence of the band edge wave-vector.
    (b) The data have been reduced to the first BZ. Continuous coloured lines represent the simulated dispersion curves for the PS, folded dispersion curves of the FS part of the sample (Rayleigh and Sezawa mode) are re-plotted as grey lines (dot-dash and dot-dot-dash respectively), black dotted (dashed) line correspond to the bulk transverse (longitudinal) mode in the substrate.}
    \label{complete_dispersion}
\end{figure*}

In the panel (a) of Fig.~\ref{data&PS//q} we report the HD-TRG data measured at
$q=0.82~\mu$m$^{-1}$ on the PS for wave-vector propagating orthogonal to the grooves. In the panel (b) the signal after the slow term subtraction is shown. Performing a FFT algorithm of the
resulting data (panel (c) of Fig.~\ref{data&PS//q}) we obtain the acoustic spectra content of
the signal, that is typically composed of more spectral contribution than the FS case.
In Fig.~\ref{complete_dispersion} the measured dispersion relation characteristic of the PS is
shown. A completely different scenario with respect to the flat region appears. On the whole, up to
four modes are detected. Even at a first look at Fig.~\ref{complete_dispersion}, the phononic
structure that elastic waves experience while propagating in the periodically corrugated surface is
clear. The bending and symmetry of the dispersion curves with respect to the BZ boundary (the band
edge at $q_{B}=\pi/5$ $\mu$m$^{-1}$) are evident. 
Two gaps are observed in SAW frequencies: one between the second and the third mode of approximately 55 MHz and the other between the third and the fourth mode of about 80 MHz. According to our experimental results, in these frequency gaps no SAW propagation is allowed, as it is for phononic band gaps. Nevertheless, SAW with different polarizations and higher wave-vectors should be investigated in order to detect whole bending and folding processes.
The first and second modes are similar to those observed in the flat region (re-plotted in
Fig.~\ref{complete_dispersion} (b) as grey lines), and in general the effect of the surface
structuring is to lower the frequency of both modes. In PS the second mode frequency is under the
transverse bulk frequency for $q \gtrsim 0.45~\mu$m$^{-1}$, whereas in the FS sample it is located
above it throughout all the investigated wave-vector range. The first mode is obviously located under
the leaky cut-off (represented by the black dotted line in Fig.~\ref{complete_dispersion}) at every
investigated wave-vector.

The third mode starts to be detected at a reduced wave-vector value of $q\simeq0.33~\mu$m$^{-1}$
with a velocity which is that of the longitudinal bulk velocity of the fused silica substrate
(dashed horizontal line). Two interpretations of the nature of this mode are possible: whether it
is a sort of \emph{optical} mode (where the word optical is used in analogy with the optical mode
of a diatomic chain) or it is an \emph{acoustic} mode (meaning that its frequency tends to zero as
the wave-vector tends to zero). The fact that we start to detect it exactly in correspondence of
the longitudinal bulk values (black dashed line in in Fig.~\ref{complete_dispersion} (b)) let us be
in favour of the latter hypothesis, since, if it would be a sort of optical mode, no restriction
would apply to its phase velocity.

Finally the fourth mode is totally different from those just discussed. It is clearly wave-vector independent as testified by its flat dispersion. The nature of this specific mode is quite obscure and could be connected with a specific vibration localized on the single stripe.

\section{DISCUSSION}

On a flat surface of an amorphous non-piezolectric semi-infinite media only a surface wave can
exist: it is named Rayleigh wave and lays in the sagittal plane~\cite{Farnell_78}. This is an
Elliptical SAW (E-SAW) composed by coupled longitudinal and shear vertical material displacements.
The Rayleigh wave is non dispersive and surface localized (i.e. its elastic energy is localized
on the matter surface since the amplitude of its displacement decay exponentially in depth of the
media, with a decay constant whose order of magnitude is that of the wavelength of the wave
itself).

The presence of a periodic corrugation or a film bounded to the substrate's surface deeply influences the propagation of SAW giving rise to new SAWs characterized by different polarizations, dispersive and localization features~\cite{Farnell_78,Gulyaev_98}. In a corrugated surface the Rayleigh wave
becomes dispersive and other E-SAW can exist~\cite{Glass_81}, besides new Shear Horizontal SAWs
(SH-SAW) can propagates on the surfaces~\cite{Auld_76}. Typically the new E-SAWs are characterized
by phase velocities higher then the transverse bulk velocity of substrate becoming ``leaky" waves
(i.e. surface waves whose elastic energy decays weakly in depth of the
media)~\cite{Giovannini_92,Lee_94}. Also the presence of a thin film on a flat surface strongly
modifies the SAWs propagation creating new surface acoustic modes~\cite{Farnell_78}. The lower
frequency E-SAW mode is again addressed to as Rayleigh mode. Besides numerous higher frequency
E-SAW modes exist and they are generally referred to as Sezawa modes, together with SH-SAWs that
are known as Love waves. Also in this case, some of these SAWs become leaky~\cite{Farnell_78}. To
our knowledge, for media presenting simultaneously corrugation and thin film neither general
theoretical nor numerical studies exist. Nevertheless the studies on the IDT, on a piezoelectric
substrate~\cite{Duhring_09} or amourphous media~\cite{Nardi_09}, present many similarities with a
thin film coated 1D PS.

In a HD-TRG experiments the signal is dominated by the transient ripple excited on the
surface~\cite{Rogers_00}. So this experiment is sensitive only to the E-SAWs that are characterized
by some vertical displacement, whereas is not sensitive to pure SH-SAW. Hence in the following
discussion we will limit to compare our HD-TRG data with the simulated E-SAW modes. In particular
some leaky E-SAW, even though elliptical, are dominated by longitudinal surface displacements
($|u|>|v|$)~\cite{Cunha_95}, so that requires an excellent signal to noise ratio in order to
experimentally detect them.

With respect to the above considerations, there are two interesting quantities to discuss: the
total displacement as function of depth in the material and a coefficient to estimate the
polarization SAW components. We define a $y$ dependent quantity that is the quadratic
\textit{total displacement} averaged along the $x$ direction,
\begin{equation}
\beta(y)=\int_{0}^{2\pi/q} [u(x,y)^2+v(x,y)^2] dx
  \label{beta}
\end{equation}
and the \textit{ellipticity-like coefficient} instead as average quadratic ratio of $u$ and $v$,
\begin{equation}
\alpha=\frac{\int\int_{0}^{2\pi/q}v(x,y)^2 dxdy}{\int\int_{0}^{2\pi/q}u(x,y)^2 dxdy}
\label{alfa}
\end{equation}

\subsection{Flat Surface}\label{Disc&hom&par}

As first step, we will analyse the HD-TRG data on the FS. In the general case of a ``slow'' film on
a ``fast'' substrate~\cite{Farnell_78}, the Rayleigh wave phase velocity $c_{R}$ is comprised
between the substrate's Rayleigh velocity $c_{Rs}$ and the film's Rayleigh velocity $c_{Rf}$ with
$c_{R}\rightarrow c_{Rs}$ for $qh\rightarrow$0, being $q$ the SAW wave-vector and $h$ the film
thickness. In our experiments the value of $qh$ remain very small for any wave-vector excited, it
varies from $qh=0.012$ to $qh=0.14$. So the phase velocity of the first mode at the lower induced
wave-vector (i.e. $qh=0.012$), is $c_{R}=3290$ m/s and it corresponds to the Rayleigh velocity of
the substrate. In fact: at this value the wavelength is much bigger than the film thickness so that
the wave does not effectively sense the presence of the film.  As much as the product $qh$ grows
the wave is more and more affected by the film presence and slows down so that the phase velocity
corresponding to the higher induced wave-vector (where $qh=0.14$) is $c_{R}=2600$m/s.

For the Sezawa waves exist a cut-off velocity for $qh\rightarrow$0 that is the transverse velocity
of the substrate $c_{Ts}$~\cite{Farnell_78}. The present measured second mode has always phase
velocities above the $c_{Ts}$, represented by the straight dotted line in
Fig.~\ref{hom&sim&par_dispersion}. So this wave is over the cut-off velocity for any wave-vectors
and it acquires a leaky nature. In order to understand the nature of these two modes we calculated
the dispersion relations of our sample by the FEA described previously. From eigenvalue analysis
only two surface modes have been retrieved. These are shown as orange and blue continuous lines in
Fig.~\ref{hom&sim&par_dispersion} and are in optimum agreement with the experimental data.
\begin{figure}
    \includegraphics[scale=0.35]{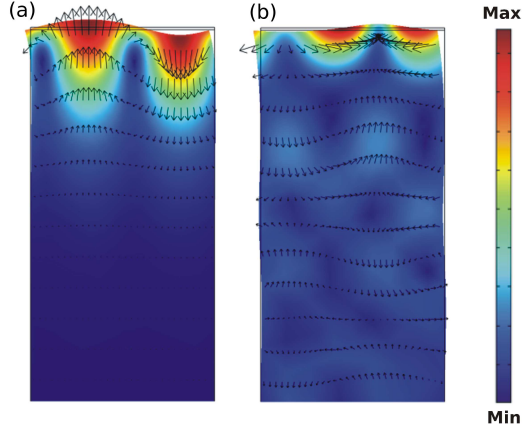}
    \caption{The simulated displacement fields for the two E-SAW modes at $q=0.628~\mu$m$^{-1}$ for the FS part of the sample. The vectorial displacements are plotted by arrows and total displacement field in colour scale: red is for maximum displacement and blue for zero motion as reported on the right side. (a) is relative to first mode (i.e. Rayleigh )and (b) to the second mode (i.e. Sezawa). The different polarization states and the different profile decays proper of the two modes are evidenced.}
    \label{FS&simul&disp}
\end{figure}
In Fig.~\ref{FS&simul&disp} we report the simulated displacement field (represented by the arrows)
and the total displacement (in colour scale) at $q=0.628~\mu$m$^{-1}$ (corresponding to a
wavelength of $10~\mu$m) of the first and second mode. Some of the general features characteristic
of the two modes throughout all the investigated wave-vector range can be appreciated. In
particular the different polarization state and the different profile decay are evident. As already
discussed there are some interesting parameters that can be obtained from the simulated
displacement fields. In panel (a) of Fig.~\ref{V&U} the wave-vector behaviour of $\alpha$ for the
second mode is shown. It can be observed that as the wave-vector increases the value of $\alpha$
also raises. This implies that as wave-vector grows the Sezawa wave acquires a vertical transverse
polarization character. Given the vertical displacement's sensitivity of the experiment, this trend
is confirmed by comparing the amplitude of the leaky SAW peak relative to $q$=0.092 $\mu$m$^{-1}$
and to $q$=1.04 $\mu$m$^{-1}$ (panel (b) of Fig.~\ref{V&U}). As expected, the intensity of this
peak (each spectra has been normalized so that the maximum of the first peak is unity) is higher
for the bigger wave-vector. Finally the different in depth decays behaviour of the two modes are
well described by the calculated $\beta$ parameter shown in panel (c) of Fig.~\ref{V&U}. As
expected the first mode decays exponentially in depth whilst the second oscillates around a non
zero value also in substrate's depth which is typical of leaky waves that involves motion also in
the bulk substrate\cite{Farnell_78}.
\begin{figure}
    \includegraphics[scale=0.5]{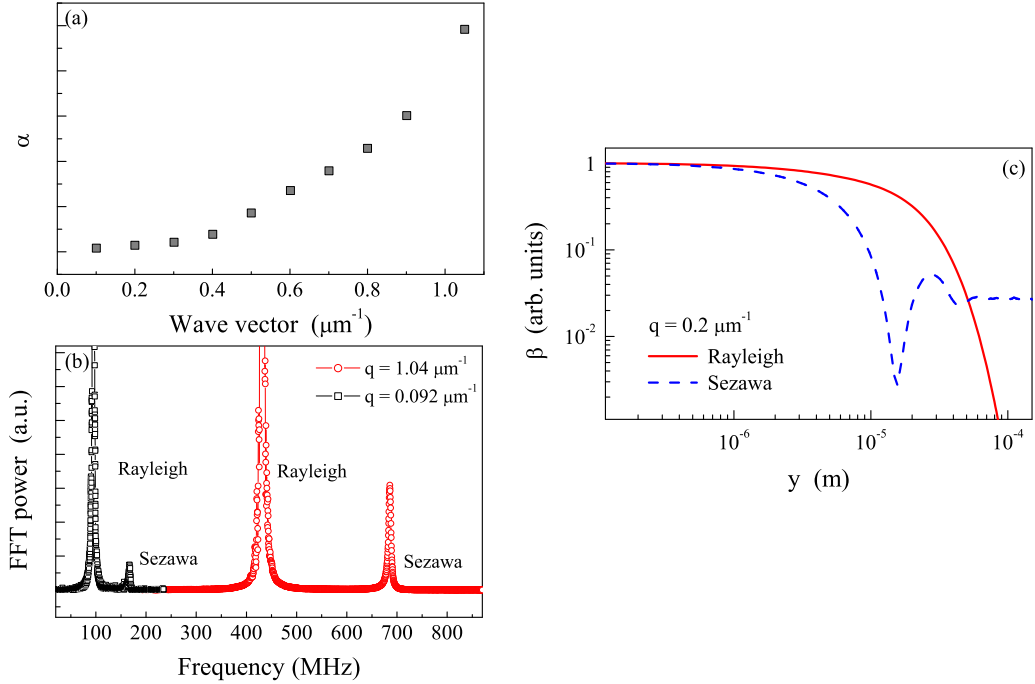}
    \caption{Analysis of the two E-SAWs in the FS part of the sample. Panel (a): ellipticity-like coefficient $\alpha$ for the second mode as a function of the wave-vector, calculated according to Eq.~\ref{alfa}. Panel (b): Power FFT of the signal at two different induced wave-vectors (each spectra has the maximum of the Rayleigh peak normalized). Panel (c): Total displacement $\beta$ as a function of depth is shown in red (blue) for the first (second) mode, calculated according to Eq.~\ref{beta}.}
    \label{V&U}
\end{figure}

\subsection{Phononic Surface}\label{orto}

The complex dispersion scenario obtained by the HD-TRG data on a PS, reported in
Fig.~\ref{complete_dispersion}, has been investigated by simulation. This has been performed using
the eigen-value analysis solved on the WSC, see Fig.~\ref{sample}(c), imposing Bloch conditions at
the boundary.  Between all the eigen-values and eingen-modes, or Bloch modes, generated by the
simulation we select that appearing in the frequency range investigated by this experiment. The
value of these eigen-values are reported as continuous lines in Fig.~\ref{complete_dispersion} (b).
As it clear they are in very good agreement with the measured dispersions. Hence, the FFT analysis
of the measured data and the simulations show that the excited SAWs can be described by 3 Bloch
modes for $q<0.28~\mu$m$^{-1}$, whereas four Bloch modes are needed for higher wave-vectors. We will
call them as first, second, third and fourth mode, starting from the lower frequency one.
\begin{figure}
\begin{center}
    \includegraphics[scale=1.3]{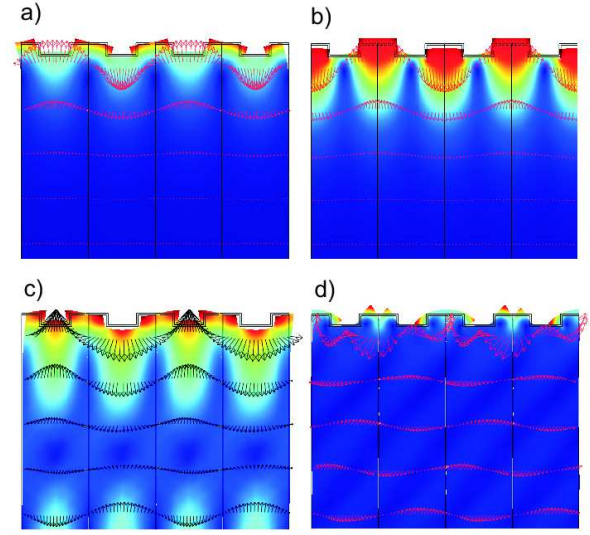}
    \caption{The simulated displacement fields associated with the 4 Bloch E-SAW modes present in the PS and corresponding to the band edge wave-vector. In the figures, the vectorial displacements plotted by arrows and total displacement field in colour scale are reported: black is for maximum displacement and white for zero motion (on the on-line colour figures they are red and blue, respectively). Plot a) is relative to first mode (i.e. lower frequency). Plot b), c) and d) are those of second, third and fourth mode respectively.}
    \label{PS&simul&disp}
\end{center}
\end{figure}
In the Fig.~\ref{PS&simul&disp} we report the displacement fields of the 4 Bloch modes observed
in the HD-TRG signal at the band edge wave-vector, $q=0.628~\mu$m$^{-1}$. These Bloch waves present
different polarization and leaky features. The first and second mode show vector and intensity
field usual of E-SAW, moreover the displacements are dominated by the vertical component.
The third mode, instead, show the features typical of a leaky-wave. Finally, the fourth mode
display a very peculiar profile that somehow differs from the normal SAWs.

We can analyse the penetration into the bulk of the different acoustic modes using the $\beta$
parameter introduced in Eq.~\ref{beta}.
\begin{figure}
\begin{center}
    \includegraphics[scale=0.4]{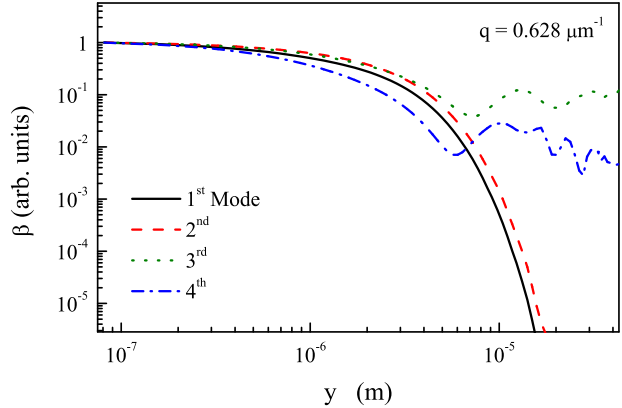}
    \caption{Total displacement, $\beta$, as a function of depth in the material,
    $y$, for the four modes present at the band edge, $q=0.628~\mu$m$^{-1}$,
    calculated according to Eq.~\ref{beta}. The first (continuous line)
    and the second (dashed line) modes  show a typical SAW decay profile.
    The third mode (dotted line) display a decay of leaky-like SAW, as well the fourth mode (dashed-dot line).}
    \label{qBE_decay}
\end{center}
\end{figure}
Fig.~\ref{qBE_decay} shows the decays of the total displacement, $\beta$, as a function of depth in
the material, $y$, obtained at the band edge, $q=0.628~\mu$m$^{-1}$. The first and the second modes
show a standard non-leaky SAW decay profile, whereas the third mode display a decay of leaky-like
SAW, as well the fourth mode. At this wave-vector both the first and the second mode are under the
cut-off frequency (transverse bulk velocity of the substrate, see Fig~\ref{complete_dispersion}):
both the decay profile and the polarization state (represented by the arrows in their relative plot
(a) and (b) in Figure~\ref{PS&simul&disp}) obtained by the simulations are in agreement with the
preceding general considerations for such waves. Third mode is located close above the transverse
cut-off and its decay profile is that typical of leaky SAW but its polarization remains dominated
in the depth by vertical polarization, whereas leaky waves are expected to acquire the longitudinal
component~\cite{Farnell_78,Giovannini_92,Cunha_95}.

\begin{figure}
\begin{center}
    \includegraphics[scale=0.35]{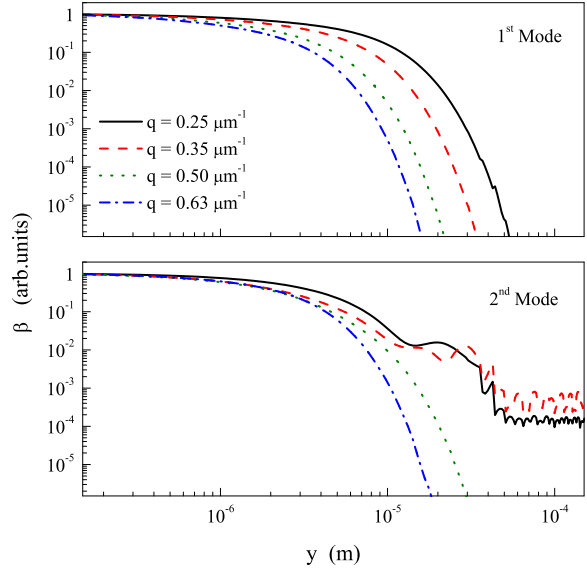}
    \caption{The total displacement decay, $\beta$, as a function of depth in the material, $y$, calculated according to Eq.~\ref{beta} for the first (top) and second (bottom) mode at four different wave-vectors: $q=0.25~\mu$m$^{-1}$ (continuous line), $q=0.35~\mu $m$^{-1}$ (dashed line), $q=0.50~\mu$m$^{-1}$ (dotted line), $q=0.628~\mu$m$^{-1}$ (dashed-dot line).}
    \label{decay}
\end{center}
\end{figure}

In Fig.~\ref{decay} it can be observed the $\beta$ parameter at four different wave-vectors, for
the two lower modes. Top graph of Fig.~\ref{decay} for the first mode at four different wave
vectors: the bigger the associated wave-vector the faster the decay, this behaviour is that typical
of surface waves since their motion is principally located within a wavelength. Bottom graph shows
the same quantity for the second mode, at the same wave-vectors ($0.25, 0.35, 0.50, 0.628$ $\mu
$m$^{-1}$). It can be observed that two decays (associated to wave-vector magnitude $0.50, 0.628$
$\mu m^{-1}$) are analogue to those of the first mode, while the others two seem to reach a plateau
value. This is associated to the fact that while the frequency associated to the two highest wave
vector is under the cut-off and consequently the decay is analogue to that of the first mode, the
two lower vectors are associated to frequencies above the cut-off and consequently they became
leaky and there is motion associated to the presence of the wave also in depth.

The fourth mode is clearly wave-vector independent as testified by its flat dispersion and its
decay, shown in Fig.~\ref{qBE_decay}, and it is much shorter than that of both leaky-SAW and SAW. Still its displacement involves also the substrate, as it can be observed both from Fig.~\ref{qBE_decay} and its displacement field, reported in the plot (d) of Figure~\ref{PS&simul&disp}. 
This mode, to our opinion, is strictly correlated to the eigen-mode of vibration of a single relief of the superficial square-wave structure. In fact, we performed a simulation of a single square-wave
relief obtaining an eigen-value very close to the fourth mode frequency. Nevertheless this
eigen-mode then slightly varies its frequency and its displacement field when the simulation is
realized over the entire sample, resulting as an eigen-mode of the whole structure.

\section{Conclusions}

In this work we have extensively characterized both from an experimental and numerical point of
view the dispersion relation of a flat and a phononic surface realized by a thin metal film on bulk
amorphous silica. We have shown that periodically micro-patterning of the surface strongly alters
the propagation of surface acoustic waves, in comparison to the flat surface, generating a complex
band structures with articulated dispersion phenomena. In particular, two new
acoustic modes of high frequency and two frequency gaps have been observed. The experimental results have been consolidated by comparison with numerical results obtained by a finite element analysis. These simulations allows to compare the measured band diagram with the calculated one but also to get
deep insight in the nature of the waves: the polarization state and surface character of the
observed waves have been discussed and compared to experimental results.

A comparison with the previously reported TRG experiments is worthwhile. Our sample is very similar
to that investigated in ref.~\cite{Dhar_00}, in fact their data are in good agreement with our
results even if in that experiment the third and fourth mode were not detected. Probably, this is due
to the lower signal-noise present in that homodyne detected TRG experiment. The papers by Maznev et
al.~\cite{Maznev_08,Maznev_09} studied a sample presenting some similarities (e.g. 1-D symmetry,
square-wave relief and micro-metric step and depth lengths) and few substantial differences
(e.g. materials composing the phononic and bulk structures) if compared with the sample here
investigated. Nevertheless, the band diagrams show some strong similarities as the clear bending of
the two lower frequency modes and a high frequency almost q-independent mode (this particularly true
for the sample investigated in ref.~\cite{Maznev_08}, whereas in the sample investigated in ref.~\cite{Maznev_09} the high frequency mode shows a q-dependence). These HD-TRG studies don't report the intermediate mode, the third one in our band diagram of Fig.~\ref{PS&simul&disp}, even if their data of Fig.3 of ref.~\cite{Maznev_08} suggest the presence of this mode. In these works the authors attempt to relate the measured band diagram to a possible general theoretical framework based on the principle of hybridization of the
acoustic normal modes, taking place if their energy/frequency get close~\cite{Giovannini_92}. In our
band diagram such a theoretical scheme does not seem to apply. According to our results the third
mode could be addressed to simply to the bending induced by the phononic structure on higher Sezawa mode. This mode could be non allowed in the FS because it would lye over the longitudinal bulk velocity limit, see
Fig.~\ref{hom&sim&par_dispersion}, but it would become possible in PS at high wave-vectors due the
bending phenomena that pushes its phase velocity below the longitudinal bulk velocity, see
Fig.~\ref{complete_dispersion}.

This experimental and numerical study shows the complex nature of the SAW propagation on PS that
results to have articulated frequency, polarization and confinement phenomena. Such peculiarity
becomes more intriguing at the BZ band edge. In our opinion, these experimental results show some interesting news that are asking for a careful comparison with the present theoretical understanding of the elastic waves propagation in phononic surfaces. The main features of three lower frequency Bloch modes, measured in this experimental investigation, can fit into the general predictions of the theoretical models~\cite{Auld_76,Glass_81,Glass_83,Giovannini_92}. Nevertheless, a specific calculation of the band diagram for our sample structure (surface grating with thin-film metal coating) would be necessary in order to complete the comparison. In particular, it would be interesting to know which folding and hybridization phenomena the theories predict for our sample, as well if they predict the existence of a $q$-independent high frequency Bloch mode.

\section{Acknowledgments}

This research has been performed at European Laboratory for Non-Linear Spectroscopy (LENS) and supported with EC grant N. RII3-CT-2003-506350. Matteo
Mannini, from LAboratory for Molecular Magnetism (LA.M.M.), Dipartimento di Chimica
Univ. di Firenze, performed the film deposition: many thanks for helpfulness and great
accuracy. Gratitude goes to Michele De Regis, a new student in our group, for the help in
collecting group velocities data. Finally we would like to acknowledge the people from CE.M.E. at
Consiglio Nazionale delle Ricerche (CNR) for the scanning electron microscope images.

\end{document}